\documentclass[12pt]{article}
\usepackage{amsmath,amssymb,graphicx,mathrsfs,slashed}
\usepackage[usenames, dvipsnames]{color}
\usepackage{tabularx}
\usepackage[%
verbose,
colorlinks=true,
naturalnames=true,
linkcolor=blue,
citecolor=black,
]{hyperref}
\usepackage{pifont}
\newcommand{\be}{\begin{equation}}
	\newcommand{\ee}{\end{equation}}
\newcommand{\bea}{\begin{eqnarray}}
	\newcommand{\eea}{\end{eqnarray}}

\def\({\left(} \def\){\right)}

\def\diag{{\text diag}}
\begin{document}
	
	\renewcommand{\baselinestretch}{1.5}\normalsize

	\title{\vspace{-1.8in}
		{ Classical Love for quantum black holes }}
	
	\author{\large Ram Brustein, Yotam Sherf
		\\
		\vspace{-.5in}  \vbox{
			\begin{center}
				$^{\textrm{\normalsize
						Department of Physics, Ben-Gurion University,
						Beer-Sheva 84105, Israel}}$
				\\ \small 
				ramyb@bgu.ac.il,\ sherfyo@post.bgu.ac.il
			\end{center}
	}}
	\renewcommand{\baselinestretch}{1.35}\normalsize
	\date{}
	\maketitle
	\begin{abstract}
		We present a method for comparing the classical and quantum calculations of the electric quadrupolar Love number $k_2$ and show that our previous derivation of the quantum Love number of a quantum black hole matches exactly the classical calculation of $k_2$ when quantum expectation values are replaced by the corresponding classical quantities, as dictated by the Bohr correspondence principle. The standard derivation of $k_2$ for classical relativistic stars relies on fixing boundary conditions on the surface of the star for the Einstein equations in the presence of an external perturbing field. An alternative method for calculating $k_2$ uses properties of the spectrum of the nonrelativistic fluid modes of the star. We adopt this alternative method and use it to derive an effective description of the interior modes in terms of a collection of driven harmonic oscillators characterized by different frequencies and amplitudes. We compare these two classical methods and find that most of the interior information can be integrated out, reducing the problem of calculating $k_2$ to fixing a single boundary condition for the perturbed Einstein equations on the surface of the deformed star. We then determine this single boundary condition in terms of the spectrum of the object and proceed to identify the relationship between classical quantities and quantum expectation values for the case of a quantum black hole and to verify the agreement between the results of the effective classical calculation and the quantum calculation.
	\end{abstract}
	\newpage
	\renewcommand{\baselinestretch}{1.5}\normalsize
	
	\renewcommand{\theequation}{\theparentequation.\arabic{equation}}
	
	\begin{subequations}
		\renewcommand{\theequation}{\theparentequation.\arabic{equation}}		
		\section{Introduction}
		
		The inspiral evolution of a binary system is accompanied by a mutual tidal interaction, leading to deformation on the spherical mass distribution.
		The tidal response of each of the companions is quantified in terms of the tidal Love numbers, which has a specific imprint on the emitted gravitational wave (GW) waveform \cite{Flanagan:2007ix,Yagi:2013baa,TheLIGOScientific:2017qsa,Abbott:2020uma,Hinderer:2007mb,Damour:2009vw,Binnington:2009bb}. In general relativity (GR), Love numbers of black holes (BHs) vanish universally as a consequence of the BH no-hair property \cite{Damour:2009vw,Gurlebeck:2015xpa,Yagi:2016bkt} (for recent discussions, see \cite{Charalambous:2021mea,Charalambous:2021kcz} and for a discussion of higher dimensional BHs, see \cite{barak}).
		
		We argued in \cite{Brustein:2020tpg} that future observations by the planned Laser Interferometer Space Antenna (LISA) \cite{LISA}  of GWs emitted during the inspiral phase of binary BH coalescence events will provide an opportunity for probing  the quantum state of macroscopic astrophysical BHs via the imprint of this state on the emitted GW. Our argument was based on showing that the electric quadrupolar Love number $k_2$, is nonvanishing for a quantum BH (QBH) and could be, under favourable circumstances, large enough for LISA to measure (similar arguments are also given in \cite{Maselli:2017cmm}). The vanishing of $k_2$ for GR BHs and it being the largest of the dimensionless Love numbers,  make $k_2$ a key diagnostic for deviations from classical GR.
		
		Modifications to classical GR could perhaps be relevant in the strong-field regime, for which quantum effects might be significant. In this regime, additional quantum scales, such as the string scale, could appear \cite{Boulware:1985wk,Zwiebach:1985uq}. Moreover, it was argued that modifications to the nature of BHs of quantum origin are needed to reconcile some fundamental problems such as the information loss paradox \cite{Braunstein:2009my,Mathur:2009hf,Brustein:2013qma}.
		Among the suggested quantum ultracompact objects that could be relevant in this context are, for example, firewalls \cite{Mathur:2009hf,Almheiri:2012rt}, fuzzballs \cite{Skenderis:2008qn}, polymer BHs \cite{Brustein:2016msz}, and many others \cite{Brustein:2020tpg,Bekenstein:1995ju, Giddings:2017mym,Cardoso:2019apo,Cardoso:2019rvt,Brustein:2019twi,Wang:2019rcf,Agullo:2020hxe}.
		
		We are interested in calculating the Love numbers of large astrophysical BHs. As for any macroscopic object,  the Bohr correspondence principle implies that some quantum state corresponds to the classical BH, no matter how large it is. We use the term ``quantum black hole" to mean the quantum state that corresponds to a classical BH. The QBH is therefore an ultracompact object that possesses a horizon and, in addition, has a discrete spectrum of quantum mechanical energy levels.
		These energy levels can be viewed as coherent states that correspond to macroscopic, semiclassical excitations of the QBH.
		For example, in the polymer BH model of a QBH \cite {Brustein:2016msz,Brustein:2019twi,Brustein:2016xzw,Brustein:2017koc,Brustein:2017nis} (see also \cite{Bekenstein:1995ju,Bekenstein:1997bt,Mathur:2005zp,Guo:2017jmi} for related models of QBHs), the interior matter can be effectively viewed as a fluid which can support pulsating modes in essentially the same way that a relativistic star does. These fluid modes would exist in addition to the standard spacetime modes of the exterior, and so their spectrum should then be added onto that of the ringdown or quasinormal modes of a perturbed BH.  Each of these fluid modes, when excited, represents a large-amplitude, high occupation number, coherent state of the interior matter rather than a single quantum excitation. In the ground state of the QBH, the exterior geometry is exactly the Schwarzschild geometry. But, when a QBH is in an excited state, it displays deviations from its GR description \cite{Brustein:2017nis}, and therefore it can be, in principle, distinguished from its classical counterpart.\footnote{In \cite{Brustein:2019twi,Kim:2020dif} it is shown that due to vacuum fluctuations, GR BHs are tidally deformed. However, the deformation leads to a nonvanishing Love number that is Planckian suppressed \cite{Kim:2020dif}.}
		This picture is consistent as long as the QBH has, to some degree, deviated from its equilibrium state, unlike a GR BH. The degree of deviation depends on the amount of energy that is injected into each of the specific modes.

		One can probe the differences between a QBH and a classical BH when they are weakly out of equilibrium in response to the external field of a binary companion. The classical BH is bald, while the quantum BH has some  quantum hair \cite{Brustein:2017nis}. Here we focus on $k_2$, a single number that is part of the quantum hair. The prevalent expectation is that quantum effects for large astrophysical BHs are controlled by the extremely small ratio of the Planck length squared to typical curvatures $l_P^2/R_S^2$, and therefore are negligibly small. However, we argued that for QBHs, quantum effects are governed by the magnitude of the quantum hair which can be much larger. For example, in string theory the dimensionless magnitude of the hair scales with the string coupling squared $g_s^2\sim l_s^2/l_P^2$, where $l_s$ is the string length. In general $g_s^2$ is expected to be small, but of the order of other typical gauge couplings $g_s^2\sim 0.1$.
		
		A classical GR BH geometry does deform when subject to an external perturbing field, however,  the deformation is not sourced by a varying matter distribution. Rather, the deformation is expressed in terms of a change in the Gaussian curvature at the Schwarzschild radius  (or, equivalently, of the scalar curvature). Then, by embedding the BH in a fictitious two-dimensional sphere, one can interpret the deformation as the relative radial deviation of the position of the Schwarzschild radius. However, this geometric deformation should not be confused with a true physical effect.   For an observer in the vicinity of the BH, the horizon is still fixed at the Schwarzschild radius. If this picture were not true, it would lead to a nonvanishing Love number. This argument is described clearly in \cite{Gurlebeck:2015xpa}, who demonstrated that geometrical deformations of classical BHs cannot be reflected in their asymptotic multipole moments and therefore their Love number vanishes identically. The conclusion is that a nonvanishing imprint on the object's asymptotic moment requires a physical matter deformation, or equivalently, a physical response of the state of the QBH to the external perturbation.

		In discussing quantum hair and its relevance for the calculation of $k_2$, we would like to highlight the fact that for an external observer, the interior of the ultracompact object is affecting the result only in terms of the one boundary condition (BC) that is imposed on the external Einstein equations at the surface of the object. The second boundary condition can be expressed solely in terms of the perturbing classical field far away from the object. In effect, the interior is integrated out of the equations and any detailed information about the object's internal composition, energy density or pressure can appear only through this one BC. The Love number is determined in terms of the ratio of the two BCs. Therefore, the only accessible information to an outside observer about the interior can be expressed in terms of the Love number(s). In this paper, we will develop the theoretical framework for applying these ideas and demonstrate explicitly how these ideas are manifested when solving the perturbed Einstein equations.
		
		The rest of the paper is organized as follows. First, we review the calculation of the quantum Love number in \cite{Brustein:2020tpg}. Then, we review the standard calculation of $k_2$ and focus on the required BCs for this calculation. In Sec.~\ref{3.2}, we review an alternative method of calculating the Love number of an ultracompact object that is driven by an external periodic force using the spectrum of its interior nonrelativistic (NR) fluid modes. We adapt this method so that the results match those for an asymptotic observer far away from the QBH. Then, we can compare the two methods and  from the consistency of the two methods and find the boundary conditions that need to be imposed on the perturbed Einstein equations. Finally, we compare our previous quantum calculation of $k_2$ to the classical calculation of $k_2$ for a QBH and show that the two calculations agree for macroscopic QBHs. This agreement, as dictated by the Bohr correspondence principle, allows us to establish the correspondence between the classical and quantum methods of calculating $k_2$ and provides us with an explicit dictionary for comparing them and showing their agreement. We end the paper with a summary and conclusion.
	\end{subequations}

	\begin{subequations}
		\renewcommand{\theequation}{\thesection.\arabic{equation}}
		\section{ Quantum Love number}\label{s22}

		We first briefly recall our previous calculation of the quantum Love number \cite{Brustein:2020tpg}, where additional details can be found.

		In GR, the interior of a BH is empty, nearly a vacuum, except for a possibly singular core. Reconciling GR with quantum mechanics led to the realization that a substantial revision to the classical GR is required. The firewall argument marked the beginning of a new era in the theory of quantum BHs \cite{Almheiri:2012rt,MP}, and forerunners of the argument, are seen in \cite{Sunny,Mathur1,Braun},\cite{Mathur2017}.
		
		We will discuss a class of solutions to the issues raised by the firewall argument. The main idea is that strong quantum effects lead to horizon scale deviations rather than  than deviations at the Planck length scale and are characterized by an intrinsic excitations spectrum. In these cases, the emergence of new physics introduces a new scale, and the ratio of the new scale to the Planck scale can be viewed as a coupling constant, as in string theory.
		The complete picture is accompanied by a self-consistent interior description that requires a significant departure from the semiclassical gravity, as well as some exotic matter which is outside the realm of the standard model \cite{BHFollies}: fuzzballs, \cite{MathurFB,otherfuzzball} and the polymer model \cite{strungout}.

		Since the interior is inaccessible to an external observer, and because the gravity in the interior is assumed to be strongly coupled, one cannot use the semiclassical geometric description in terms of a curved spacetime.  However, the only relevant aspect of the QBH interior is that excitations are macroscopic so applying Bohr correspondence principle is justified, implying that the excited spectrum of a QBH can be described by a set of coherent states. Then, from the Bohr correspondence principle, we conclude that these states correspond to semiclassical states that can be effectively described as an oscillating classical system. These assumptions were previously discussed also in \cite{Bekenstein:1997bt,Bekenstein:2020mmb} and later in \cite{Hod:1998vk,Maggiore:2007nq}. In these works, the macroscopic excitations and the energy spectrum of a quantized BH were described by a spectrum of a harmonic oscillator.
		
		Then, to implement the idea, we view the exotic matter in the interior of the QBH effectively as a fluid that supports pulsating modes as for a relativistic star. These fluid modes would exist in addition to the standard spacetime modes of the exterior.  The  perturbations are divided into two sectors, the fluid modes and spacetime modes. Due to their low speed of sound and the compactness of the QBH, fluid modes are decoupled from the spacetime perturbations as in the  Cowling approximation \cite{Kokkotas:1999bd,Allen:1997xj,Andersson:1996ua}. The fluid mode description is discussed  extensively in Sec.~\ref{3.2}.
		
		As a motivation for the	calculation of the Love number of the QBH, let us recall the analogous calculation of the polarizability of an atom. Consider an atom in its ground state $|\Psi_0\rangle$, which is labeled by the quantum numbers $|n,l,m \rangle=|1,0,0\rangle$ and assume that the expectation value of the quantum dipole operator $\widehat{D}_i$, vanishes in the ground state. Recall that, classically, the dipole moment is given by $\vec{D}=\int\rho(\vec{x}')\vec{x}'dV'=0$, where $\rho$ is the charge density of the atom. The atom is placed in a region of an approximately uniform electric field $\mathcal{E}_i$ that is induced by a weak external potential $U_{ext}$, $\mathcal{E}_i=\frac{\partial U_{ext}}{\partial x^i}$.  The interaction of the atom with the external electric field $\widehat{V}_{int}$, is expressed in terms of  $\widehat{D}_i$,  $\widehat{V}_{int}~=~-\mathcal{E}_i\widehat{D}_i$.
		The induced dipole moment of the perturbed atom in second-order time-independent perturbation theory, is given by the standard textbook expression,
		\begin{gather}
			\langle\Psi_0|\widehat{D}_j|\Psi\rangle~=-\mathcal{E}_i\sum_{n\ne 1,l,m} \dfrac{\langle 1,0,0|\widehat{D}_i|n,l,m\rangle\langle n,l,m|\widehat{D}_j|1,0,0\rangle }{\Delta E_{1,n}}~,
			\label{idm}
		\end{gather}
		where  $\Delta E_{1,n}= E_1-E_n$. In this case, symmetry implies that $l=1$, $m=-1,0,1$ and $i=j$.
		The atom's linear response to the external electric field is then $\langle\Psi_0|\widehat{D}_i|\Psi_0\rangle=\alpha \mathcal{E}_i$, where $\alpha$ is the electric polarizability,
		\begin{gather}
			\alpha~=~\sum_{n\ne 1,m+-1,0,1} \dfrac{|\langle 1,0,0|\widehat{D}_i|n,1,m\rangle|^2 }{\Delta E_{1,n}}~.
		\end{gather}
		
		Following similar considerations, we derived in \cite{Brustein:2020tpg} an expression for the gravitational polarizability, the Love numbers. The idea is to replace the electric field and the dipole moment by the tidal fields and the mass moment.

		For the quantum calculation, we considered, again, the inspiral phase of a binary system, as we did for the classical calculation. One of the companions is an object of mass $M_{ext}$ on a circular orbit of radius $b$ and the other is a  QBH of mass $M_{BH}$ and radius $R_S= 2 M_{BH}$.  In the early stages of the inspiral, the QBH responds to the external slowly varying tidal field that is generated by its companion. For $b\gg R_S$ one can expand the Newtonian potential $U_{ext}~=-{M_{\text{ext}}}/{|\vec{b}-\vec{x}|}$, of the external body in the vicinity of the QBH in its local inertial frame,
		$
		U(t,x)_{ext}~=~U_{ext}(0)+\frac{1}{2}\frac{\partial^2 U_{ext}}{\partial x^i \partial x^j}\big|_0x^{i'}x^{j'} +\cdots.
		$
		
		As in the case of the electric polarizability, the interaction of the QBH with the external field is expressed in terms of the mass moment expectation value, $\widehat{Q}^{(l)}$. These operators are the quantum counterparts of the classical symmetric trace-free mass multipoles \cite{Thorne:1980ru}. We further assume that the expectation value of the BH mass moment vanishes in the BH ground state, as dictated by the spherical symmetry and the classical no-hair properties. Denoting the ground state of the QBH by $|\Psi_0\rangle $,
		we have $\langle\Psi_0|\widehat{Q}^{(l)}|\Psi_0\rangle=0$.
		Owing to the slowly varying weak external potential, time-independent perturbation theory is a good approximation.
		
		In \cite{Brustein:2020tpg} we evaluated the correction to the ground state energy due to the induced quadrupole, $\widehat{Q}_{ij}$ of a nonrotating quantum BH whose Schwarzschild radius is $R_S$.  Recall that in the classical case, ${Q}_{ij}=\int {\rho}(t,x^{'}) \left(x^{'}_ix^{'}_j-\frac{1}{3}r^{'2}\delta_{ij}\right) dV'$, where ${\rho}$ is the energy density. In an analogy to the electric polarizability calculation, the interaction energy is given by
		$\widehat{V}_{int}=-\tfrac{1}{2}  \mathcal{E}_{ij}\widehat{Q}_{ij}$,
		where $\mathcal{E}_{ij}~=\frac{\partial^2 U_{ext}}{\partial x^i \partial x^j}$.

		The leading order corrections to the QBH ground state quadrupole in second-order time-independent perturbation theory is given by
		\begin{gather}
			\langle\Psi_0|\widehat{Q}_{kl}|\Psi_0\rangle~=~ \mathcal{E}_{ij}\sum_{n> 1,l,m} \dfrac{\langle \Psi_0|\widehat{Q}_{ij}|n,l,m\rangle \langle n,l,m|\widehat{Q}_{kl}|\Psi_0\rangle}{|\Delta E_{1,n}|}~,
			\label{qi}
		\end{gather}
		where $|\Delta E_{1,n}|= E_n-E_1$. Here the radial number of the ground state $\Psi_0$ is denoted by $n=1$, so the energy of the ground state is $E_1=M_{BH}$. This choice is made for consistency with the standard treatments of second-order perturbation theory.
		The electric quadrupolar Love number is defined as the proportionality coefficient between the induced electric quadruple moment to the external tidal field
		\begin{gather}
			\langle\Psi_0|\widehat{Q}_{ij}|\Psi_0\rangle=-\lambda_2 \mathcal{E}_{ij}~.
			\label{lqm}
		\end{gather}
		Here $\lambda_2$ is the dimensional quadrupolar Love number, which in its dimensionless form is commonly defined as  $k_2=\frac{3}{2}R^{-5}\lambda_2$. From Eq.~(\ref{qi}) it follows that	
		\begin{gather}
			k_2~=~-\dfrac{3}{4 R^{5}}\sum_{n,-2<m<2} \dfrac{|\langle \Psi_0|\widehat{Q}_{ij}|n,2,m\rangle|^2 }{|\Delta E_{1,n}|}~.
			\label{k22}
		\end{gather}
		
		We would like to point out that the additional QBH excitations decay at a parametrically slow rate in comparison to the Schwarzschild time due to the large redshift factor in the vicinity of the deformed QBH.  This was explained in detail in \cite{Brustein:2017nis} and we recall the essence of relevant arguments below. The excited modes therefore have a parametrically small width compared to the standard GR BH excitations, which justifies neglecting the width of the excitations in Eq.~(\ref{k22}).
		
		For an exterior observer,  the wavelength of the excited modes near the source  must be $\lambda_S\sim R_S$, which then asymptotically redshifts to some larger value, $\lambda_A\sim R_S z$ with $z\gg 1$. This observer then assigns a transmission cross section for such long wavelength modes through a proportionally smaller surface of area $A$ which is determined by the ratio $A/\lambda^2\sim 1/z^2$. The coupling or efficiency of emission then scales as $1/z^2$. The damping time for that mode $\tau$ is related to the inverse of the efficiency of emission and therefore scales as $z^2\gg 1$.

	\end{subequations}

	\begin{subequations}
		\renewcommand{\theequation}{\thesection.\arabic{equation}}
		\section{Classical Love number}
		
		\subsection{Method 1: Explicit internal solution for the metric}
		
		We review here the standard calculation of the Love number. This will also be used to set up notations and conventions.
		
		The response of a body to a weak external tidal field is reflected in its induced mass (electric) and current (magnetic) moments.  We will focus here on the quadrupolar electric Love number $k_2$. At large distances, in the star's local asymptotic rest frame, the temporal component of the metric is given by
		\begin{eqnarray}
			g_{tt}&=&   -1+\dfrac{2M}{r}-\mathcal{E}_{ij}x^i x^j +3\frac{1}{ r^5}Q_{ij}x^ix^j \cr
			&=& -1+\dfrac{2M}{r}-\mathcal{E}_{ij}x^i x^j-2k_2 (R/r)^5\mathcal{E}_{ij} x^i x^j
			\label{gtt},
		\end{eqnarray}
		where $M,R$ are the mass and the radius of the star, respectively and $k_2$ is the dimensionless tidal Love number that measures the linear response to the applied field.
		The first term in the deviation of $g_{tt}$ from the Schwarzschild metric in Eq.~(\ref{gtt}) describes the applied tidal field, while the term proportional to $k_2$ describes the induced trace-free quadrupole moment $Q_{ij}=\int d^3x \rho(x)(x_ix_j-\frac{1}{3}\delta_{ij}r^2)$,
		\begin{gather}
			Q_{ij}~=~-\dfrac{2}{3}k_2R^5 \mathcal{E}_{ij}.
		\end{gather}
		From a Newtonian perspective, at large distances $g_{tt}=-(1+2U_{N})$. The expansion of the Newtonian potential $U_N$, to second-order in the body's local inertial frame reads $U_{N}=-\frac{M}{r}-\frac{3}{2 r^5}Q_{ij}x^ix^j+\frac{1}{2}\mathcal{E}_{ij}x^ix^j$, with $\mathcal{E}_{ij}$ being the quadrupole moment of the external potential $\mathcal{E}_{ij}=\frac{\partial^2 U_{ext}}{\partial x_i x_j}$. Here we discuss the axisymmetric external potential, so the tidal field is given by $\mathcal{E}_{ij} x^i x^j = \mathcal{E} r^2 Y_{20}$ and consequently also the induced moment has the same angular dependence $Q_{ij}=Q Y_{20}$. It follows that
		\be
		U_N= -\dfrac{M}{r} -\dfrac{3}{2 r^3} Q Y_{20} + \frac{1}{2} \mathcal{E} r^2 Y_{20}
		\ee
		and
		\be
		\begin{aligned}
			g_{tt} &=
			-1+\frac{2M}{r}  + 3 Q \dfrac{1}{r^3}Y_{20} - \mathcal{E} r^2 Y_{20}
			\\[10pt]
			&= -1+\frac{2M}{r}  - 2 k_2 R^5 \mathcal{E} \dfrac{1}{r^3}Y_{20} - \mathcal{E} r^2 Y_{20}
		\end{aligned}
		\label{gtt1}
		\ee
		with
		\be
		k_2 R^5=- \frac{3}{2} \frac{Q}{\mathcal{E}}.
		\label{k2def}
		\ee

		We now turn to the review of the calculation of the Love number $k_2$ \cite{Hinderer:2007mb,Damour:2009vw,Binnington:2009bb}.
		One considers a perturbation $h_{\mu\nu}$  about the Schwarzschild background $g^{(0)}_{\mu\nu}$,  $g^{(0)}_{\mu\nu}~=~\diag\left(-e ^{\nu(r)},e^{\lambda(r)},r^2,r^2 \sin^2 \theta\right)$, with $e ^{\nu(r)}=e^{-\lambda(r)}=1-2M/r$.
		Then,  $h_{\mu\nu}$ is decomposed into even-parity and odd-parity parts as in the Regge-Wheeler gauge. Focusing on the the $l=2$, $m=0$ term, to linear order, in the limit that the external field is static  $\mathcal{E} =const.$, the even-parity perturbations can be expressed as \cite{Regge:1957td}
		\begin{gather}
			h_{\mu\nu}~=~\diag\left(e ^{\nu(r)}H_0(r),e^{\lambda(r)}H_2(r),r^2K(r),r^2 \sin^2 \theta K(r)\right)Y_{20}.
			\label{perteq1}	
		\end{gather}
		Then, solving the perturbed Einstein equations outside the body, one finds that $H_0=H_2\equiv H(r)$ and arrives at the following perturbation equations
		\begin{align}
			&H''+\dfrac{2x}{x^2-1}H'-\dfrac{6 x^2-2}{(x^2-1)^2}H~=~0~,
			\label{Hpp}\\
			&K'-H'-\dfrac{2}{(x^2-1)}H~=~0~,
			\label{Kprime}\\
			&K-\dfrac{1}{2}H'-\dfrac{1}{4}\left(\dfrac{x+1}{x-1}+\dfrac{3x+5}{x+1}\right)H~=~0~. \label{K2}
		\end{align}
		where $x=r/M-1$ and the prime denotes a derivative with respect to $x$.

		The exterior solution of the perturbation equations is given by,
		\begin{equation}
			H_{\text {ext}}(x)={c_1} \left[\frac{ x \left(5-3 x^2\right)}{x^2-1}+\frac{3}{2}  \left(x^2-1\right) \ln \left(\frac{x+1}{x-1}\right)\right]+3 {c_2} \left(x^2-1\right),
			\label{Hex}
		\end{equation}
		\be
		K_{\text ext}(x)= -c_1 \dfrac{4+3x(x+3)}{x+1} +\dfrac{3}{2} c_1 (x^2+2 x -1) \ln\left(\frac{x+1}{x-1} \right) + 3 c_2 (x^2+2x-1)
		\label{kex}
		\ee
		and the perturbed background metric by
		\be
		g_{tt}=-\left(\dfrac{x-1}{x+1}\right)\left(1+H_{\text {ext}}Y_{20}\right).
		\label{gb}
		\ee
		We will be  interested in cases for which $x-1= r/M-2\ll 1$. For later use, we expand $H_{\text {ext}}$ and $K_{\text {ext}}$ in this limit,
		\begin{equation}
			H_{\text {ext}}(x)= \frac{c_1}{ x-1}-3c_1 \ln \left(\frac{x-1}{2}\right)+6 {c_2} (x-1) + \mathcal{O}(x-1),
			\label{Hexex}
		\end{equation}
		\be
		K_{\text {ext}}(x)= -8 c_1 -3~ c_1 \ln\left(\frac{x-1}{2}\right) + 6 ~c_2 + \mathcal{O}(x-1).
		\label{kexex}
		\ee

		It is convenient to relate the coefficients of the expanded metric in Eq.~(\ref{gtt1}), $\mathcal{E}$ and $k_2$, to the coefficients of the expanded metric in Eq.~(\ref{Hex}), $c_1$ and $c_2$:
		\be
		c_1=40 M^2 k_2 \mathcal{E}
		\label{c1k2E}
		\ee
		and
		\be
		c_2=\dfrac{1}{3} M^2 \mathcal{E},
		\label{c2E}
		\ee
		so
		\begin{equation}
			k_2= \dfrac{1}{120} \dfrac{c_1}{c_2}.
			\label{k2c1c2}
		\end{equation}

		Before proceeding to discuss explicit solutions for $k_2$, let us emphasize some features about its dependence on the deformed body. First, we note that the functional form of the exterior solution Eq.~(\ref{Hex}) does not depend on the interior, since it corresponds to a vacuum solution of the Einstein equations. Because $c_2$ is determined entirely by the external field, the only implicit dependence of $H_{\text {ext}}$ on the interior is encoded in the ratio $c_1/c_2$ or, equivalently, the magnitude of $k_2$. It follows that one needs to specify one more BC on $H_{\text {ext}}$ to determine completely the form of the external metric perturbations.

		The standard approach for calculating $k_2$ for objects whose interiors are known and well-defined is reviewed in \cite{Hinderer:2007mb,Damour:2009vw}. One starts the calculation by specifying the interior configuration of the perturbed body in terms of its stress-energy-momentum tensor and an equation of state. Then, the perturbed Einstein equations in the interior, which have the same form as the exterior equations~(\ref{Hpp})--(\ref{K2}), are solved and their solution  $H_{\text {int}}$ is found. Next, demanding regularity of $H_{\text{int}}$ at the star's center $r=0$ and continuity of $H$  and $K$ at the star's surface $r=R$, one finds the coefficients $c_1$ and $c_2$, or equivalently $k_2$.
		
		It is clear that this method of calculating $k_2$, in which one needs the full detailed solution of the interior, contains a lot of redundant information if one is just interested in finding one number  -- $k_2$. We just need one ratio of two numbers for that, so any other method for specifying this ratio would work just as well as the standard one. We will describe such an alternative method in the next section.
		
		\subsection{Method 2: Spectrum of nonrelativistic fluid modes } \label{3.2}

		In this section we discuss a binary system, in which the object of interest -- the ``primary" -- is driven by a weak periodic force which is exerted by the companion. We rely on  the ideas presented in \cite{Lai:1993di,Ho:1998hq,Chakrabarti:2013xza,Andersson:2019ahb} to establish an effective description for the interior fluid modes of ultracompact objects as a collection of driven harmonic oscillators characterized  by their frequencies.
		In this effective description, the interior modes of the object are described from the point of view of an asymptotic observer as NR fluid modes and are analyzed in a similar manner to the analysis of classical Newtonian NR fluid modes.  As discussed in Sec.\ref{s22},  due to their low speed of sound and the compactness of the QBH, fluid modes are decoupled from the spacetime perturbations as in the  Cowling approximation \cite{Kokkotas:1999bd,Allen:1997xj,Andersson:1996ua}. This splits the interior perturbation into two sectors, fluid modes, which we will discuss, and spacetime modes, which we will neglect.
		
		We consider the oscillating modes of the object that are labeled by the radial and spherical indices $n,l,m$. Again, we focus on the tidal axisymmetric perturbations, so $l=2$, $m=0$.
		The analysis proceeds by writing and solving the equations for the Lagrangian displacement vector of the fluid $\xi^i$ which is proportional to $Y_{20}$.  The total displacement vector is given by the complete sum over the contributions from each of the radial modes,
		\begin{gather}
			\xi^i ~=~\sum_n a_n \xi^i_n.
		\end{gather}
		The modes $\xi_n$ have units of length, so the coefficients $a_n$ are dimensionless. Different radial modes are orthogonal and their precise normalization will not be important for us.
		
		In the absence of the driving force that arises from the binary companion, the fluid modes satisfy the following Harmonic-oscillator equation:
		\begin{gather}
			\ddot{a}_n +\omega_n^2 a_n~=0,
		\end{gather}
		with $\omega_n$ being the frequency of the $n$'th mode.
		When an external tidal potential is present, the equation of motion (EOM) for the internal fluid modes becomes that of a driven harmonic oscillator (see, for example, \cite{Lai:1993di}),
		\begin{gather}
			(-\omega^2 +\omega_n^2)a_n =~\frac{\mathcal{E} Q_n}{MR^2}.
			\label{DHO}	
		\end{gather}
		Here $\omega$ is the frequency of the external tidal field, to be discussed shortly.
		The quadrupole associated with the $n$th mode $Q_n$, is defined by the overlap integral
		\be
		Q_n=-\int d^3 r \delta \rho_n r^2.
		\label{oi}
		\ee
		The quadrupolar energy density perturbation $\delta \rho_n$, is associated with the $n$th fluid mode and $\Delta E_n$ is the corresponding total mass quadrupole moment.
		$
		Q_n= -\gamma \Delta E_n R^2
		$,
		with $\gamma$ being a dimensionless number of order unity.
		
		We are interested in the $m=0$ modes, so the driving is essentially at zero frequency, with $\omega=m\Omega$, $\Omega=\sqrt{M/b^3}$ being the orbital frequency. So, Eq.~(\ref{DHO}) simplifies,
		\begin{gather}
			\omega_n^2 a_n~=\frac{\mathcal{E} Q_n}{MR^2}.
			\label{DHOsol}
		\end{gather}
		In general, the driving frequency can be neglected  also for the case $m\neq 0$ as it is small compared to the oscillator natural frequencies $\Omega^2\ll \omega^2_n$,  $\omega_n^2 \sim 1/R^2$, while $\Omega^2= R/b^3$ and $R^3/b^3\ll 1$.
		The solution of Eq.~(\ref{DHOsol}) is the following,
		\be
		a_n=\mathcal{E} \dfrac{Q_n}{M \omega_n^2 R^2}.
		\label{ansol}
		\ee
		
		Next, we identify the Love number using the asymptotic moments for a static observer at infinity which can be read off from the external metric perturbation equation~(\ref{k2def}), with $Q=\sum\limits_na_n Q_n$ \cite{Chakrabarti:2013xza}, in contrast to \cite{Andersson:2019ahb} where the Love number is identified at the star surface.
		Substituting $\mathcal{E}$ from Eq.~(\ref{ansol}) into Eq.~(\ref{k2def}), one finds
		\begin{gather}
			k_{2n} R^5~=~-\dfrac{3}{2}\dfrac{Q_n}{\mathcal{E}} ~=~-\dfrac{3}{2}\dfrac{1}{a_n}\dfrac{Q_n^2}{M\omega_n^2 R^2}.
		\end{gather}
		Now, from the asymptotic moment decomposition $k_2=\sum\limits_n a_n k_{2n}$, we obtain
		\begin{gather}
			k_{2}~=~-\sum_n \dfrac{3}{2R^5}\dfrac{Q_n^2}{M\omega_n^2R^2}.
			\label{k2sum}
		\end{gather}
		We can re express the results in terms of the the intrinsic energy spectrum of the driven system. We may define the intrinsic energy difference $\Delta E_n^{\text{int}}=E_n^{\text {int}}-M$ as
		\be
		\Delta E_n^{\text {int}}=\tfrac{1}{2}M\omega_n^2 R^2.
		\label{DEint}
		\ee
		The intrinsic energy difference $\Delta E_n^{\text {int}}$ depends only on the intrinsic properties of the object and does not depend on the external driving field. It should not be confused with the energy that is pumped into the mode $n$ by the external field,
		\be
		\Delta E_n^{\text {induced}}= \tfrac{1}{2} M R^2 \omega_n^2  a_n^2.
		\label{DEind}
		\ee
		Substituting $a_n$ from Eq.~(\ref{ansol}) into Eq.~(\ref{DEind}),  we find that $\Delta E_n^{\text {induced}}$ is related to the total work done by the external tidal force,
		\be
		\Delta E^{\text{induced}}~=~\sum_n\Delta E_n^{\text {induced}} = \tfrac{1}{2} \sum_n \mathcal{E} a_nQ_n=\tfrac{1}{2}\mathcal{E} Q.
		\ee

		Substituting Eq.~(\ref{DEint}) into Eq.~(\ref{k2sum}),
		we obtain our final expression for the Love number,
		\begin{gather}
			k_{2}~=~-\dfrac{3}{4 R^5}\sum_{n} \dfrac{Q_n^2}{\Delta E_n^{\text {int}}}.
			\label{k2fluid}
		\end{gather}
		In the case that the sum is dominated by the lowest energy level $n=1$, then,
		\be
		k_{2}~\simeq~-\dfrac{3}{4 R^5}\dfrac{Q_1^2}{\Delta E_1^{\text {int}}}.
		\label{estk2fluid}
		\ee
		
		We can further parametrize $Q_n$ on dimensional grounds, as was previously done, by
		\be
		Q_n= \gamma_n \Delta E_n^{\text {int}} R^2,
		\label{qngamma}
		\ee
		where $\gamma_n$ is a dimensionless number that depends on the detailed functional form of the energy density profile of the object.
		Then,
		\be
		k_{2}~=~-\dfrac{3}{4 R} \sum_n \gamma_n^2 \Delta E_n^{\text {int}}.
		\label{k2fluidapp}
		\ee
		On general grounds, we expect $\gamma_n$ to rapidly decrease as $n$ increases, as we expect the higher-$n$ modes to possess more nodes and thus induce a smaller quadrupole moment.\footnote{This property is  typical for driven harmonic systems, as suggested by the equation~(\ref{DHO}).}
		If the decrease in $\gamma_n$ offsets the increase in $E_n$ as expected,
		\be
		k_{2}~\simeq~-\dfrac{3}{4 R} \gamma_1^2 \Delta E_1^{\text {int}}.
		\label{k2fluidapp1}
		\ee
		This relation can be further simplified since  $\Delta E_1^{\text{int}}\sim M \omega_1^2 R^2$ and typically $\omega_1\sim 1/R$, so $k_2\sim \gamma_1^2$.
		
		In summary, the calculation of the Love number amounts to a classical linear response calculation of a collection of driven harmonic oscillators by an external force in the limit that the intrinsic frequency of the oscillator is much higher than the frequency of the driving.
		
		In the quantum case, the calculation is similar, except that the relevant quantity is the quadrupole moment of the interior modes (evaluated by a static observer at infinity) without an explicit reference to the Lagrangian displacement vector or to the Newtonian potential at the surface of the star.  We emphasize that these should be viewed as means to an end: fixing one additional BC for the exterior perturbation equations. This will be our bridge to the quantum Love calculations and results in the semiclassical approximation. We need to know the energy of the lowest lying level and the quadrupole associated with this level.
		
		\subsection{Relating the two methods for quantum black holes }	
		

		The two methods of calculating the Love number calculate the same quantity, the response of an ultra compact object to an external perturbation and so, in principle, must agree. The only issue is to what extent the various approximations that are made along the way affect the resulting values of $k_2$ that are obtained by the two methods.
		
		However, we now argue  that a quantitative relation between the fluid and geometric methods is not necessary for our purposes, as we are interested in a general relationship that can then be applied to QBHs. We emphasize that for the QBH it is also impossible to find directly such a quantitative relation without a more detailed description of the interior. We bypass the fact that the interior of the QBH is not prescribed by observing that for QBHs all that is required to determine the Love number is a single BC at the surface of the QBH.
		
		The physical assumptions that we make to identify this single BC are the following:
		\begin{enumerate}
			\item
			Both GR BHs and QBHs possesses a horizon.
			\item
			In the absence of the external perturbation, QBHs cannot be distinguished from GR BHs.
			\item
			For QBHs, geometric deformations induced by external perturbations are reflected in their asymptotic moments, in contrast to GR BHs. This implies that,  due to changes in their state, the QBHs possess hair that leads to a nonvanishing $k_2$.
			\item
			For QBHs, as for their classical counterparts, the external metric perturbation must vanish on the deformed horizon.
			\item
			The interior excited modes of the QBH can be described effectively as a collection of driven harmonic oscillators, which gives rise to the Love number as in Eq.~(3.25). An exact solution of $Q_n$  as in Eq~(3.21) is model dependent and can be parametrized by a dimensionless number of order unity $\gamma$ as in Eq~(3.31).
		\end{enumerate}

		To apply our ideas in a concrete context, let us consider the temporal component of the metric perturbation $\delta g_{tt}$ near the boundary of the QBH. For simplicity, we suppress the angular dependence of the external metric perturbation. Denoting by $x_B$ the value of $x$ at the deformed surface, we get
		\begin{gather}
			-\delta g_{tt}(x_B)~\sim ~c_1+c_2(x_B-1)^2.
			\label{dgtt}
		\end{gather}
		We assume that $x_B-1\ll 1$, so denoting by $\Delta R$ the difference $R-2M$, $\Delta R /2M\ll 1$, it follows that
		\be
		x_B-1 ~=~  \frac{\Delta R}{M}.
		\ee
		We need to choose a single additional BC that the classical metric external to the QBHs needs to satisfy, that is, we need to choose one more BC that $H_{\text {ext}}$ needs to satisfy at the deformed boundary of the QBH.
		Applying our five assumptions above, we find that the following conditions need to be imposed:
		\begin{enumerate}
			\item
			Assumption 1 implies that for both classical and quantum BHs, for $\mathcal{E}=0$ (or equivalently, $c_2=0$), $x^{_{BH}}_B=x_B^{_{QBH}}=1$ and $g_{tt}(x_B)=0$, in agreement with assumption 4.
			\item
			Assumption 2 implies that  both GR BHs and QBHs possess a horizon and
			cannot be distinguished in the absence of perturbations. Furthermore, according to assumption 3, for $\mathcal{E}\ne 0$  a physical deformation of the QBH is induced, such that the surface of the QBH is shifted to $x_B=1+\delta x_B$. The specific value of $\delta x_B$ depends on the spectrum of the QBH.
			\item
			Assumption 1 implies that for classical BHs $c_1=0$ and $x_B^{_{BH}}=1$ so $\delta g^{_{BH}}_{tt}\sim c_2 (x_B-1)^2=0$ and from Eq.~(\ref{gb}),  $g^{_{BH}}_{tt}(x_B^{_{BH}})=0$. Assumption 4 implies that the QBH case is similar,  $ g_{tt}(1+\delta x_B)=0$ where $x_B=1+\delta x_B$ is the position of the deformed horizon.
		\end{enumerate}
		These three  conditions are summarized in Table \ref{tb1}.
		\begin{table}
			\begin{center}	
				\begin{tabular}{c|c|c|c|c}
					&$x_{B}$& $\delta g_{tt}$       &$\Delta R/R$  &$k_2$   \\ \hline
					QBH  &$1+ \delta x_B$ & $ g_{tt}(1+ \delta x_B)=0 $ &$c_2-c_1$ &   $\ne 0$
					\\ \hline \cline{1-3}
					BH & 1& $\delta g_{tt}(x_B)=0 $         &   $c_2$ &0
				\end{tabular}
				\caption{Comparison of the response of classical and quantum BHs. We evaluate the Euclidean deformation $\Delta R/R$ from the Gaussian curvature Eq.~(\ref{kex}) at the surface, $K(x_B)\sim \Delta R/R$, \cite{Damour:2009vw}.}
				\label{tb1}
			\end{center}
		\end{table}
		
		We wish to emphasize the origin for the difference between GR BHs to QBHs, where according to \cite{Damour:2009vw,Binnington:2009bb} it is the boundary condition on the BH horizon that kills the response terms [the terms proportional to $c_1$ in Eq.~(\ref{Hexex})] and leads to the vanishing of the Love number. An observer in the vicinity of a  classical BH sees no deviation in the horizon
		position; the horizon is ’frozen’ at $ R = 2M $, and the external perturbation is \textit{singular} on it. On the other hand, according to assumptions 3 and 4, the horizon of a QBH does deform (see discussion in Sec.~\ref{3.2}). Thus, the boundary conditions on the deformed surface are \textit{regular} and lead to a nonvanishing Love number.
		
		Rather than imposing $ g_{tt}(1+ \delta x_B)=0$, we impose an equivalent condition,
		$\delta g_{tt} (1) = -\frac{1}{2} c_1 Y_{20}$, as we now explain. The fact that some points with $x_B=1$ are formally within the original horizon is not relevant for our discussion, as we are only interested in the perturbed metric far away from the horizon of the QBH. The classical metric is, of course, only valid outside the QBH horizon.
		
		Let us recall the expansion of $H_{\text {ext}}$  in Eq.~(\ref{Hexex}),
		\be
		H_{\text {ext}}(x_B) = \frac{c_1}{(x_B-1)} + \mathcal{O}(x_B-1)=120  k_2\frac{c_2}{x_B-1}+ \mathcal{O}(x_B-1),
		\label{He1}
		\ee
		where we used Eq.~(\ref{k2c1c2}) to relate $c_1$ to $c_2$.
		It follows that
		\be
		\delta g_{tt}(x_B)= -\tfrac{1}{2} c_1   + \mathcal{O}(x_B-1) =5 k_2 R^2 \mathcal{E}+  \mathcal{O}(x_B-1),
		\ee
		where we used  the relationship between $c_2$ and  $\mathcal{E}$ in Eq.~(\ref{c2E}).
		
		The choice of BC for fixing the exterior solution is clear. Since $\mathcal{E}$ (or equivalently, $c_2$) is already fixed by the BC at infinity, we need to fix $\delta g_{tt}(x_B)$ to fix the value of $c_1$. Here we use our assumptions 3 and 5 to choose the value of $c_1$  such that the value of $k_2$ agrees with the value obtained using the fluid calculation in Eq.~(\ref{k2fluid}),
		\be
		c_1 = -2 \delta g_{tt}(2M)= \frac{15}{2}\dfrac{\mathcal{E}}{ (2M)^3} \sum_n \dfrac{Q_n^2}{\Delta E_n^{\text {int}}}.
		\label{finalc1}
		\ee
		
		Equation~(\ref{finalc1}) completes our comparison between the two classical  methods of calculating $k_2$ as it determines the relationship between the calculation of $k_2$ in terms of the spectrum of the  fluid modes  and the choice of BC on the perturbed relativistic Einstein equations for the case of a QBH.

		If the lowest level dominates the sum, as expected, we can use the estimate in Eq.~(\ref{estk2fluid}),
		\be
		c_1=-2 \delta g_{tt}(R_B)\sim- \frac{15}{2}\dfrac{\mathcal{E}}{ (2M)^3} \dfrac{Q_1^2}{\Delta E_1^{\text {int}}}.
		\ee
		Furthermore, from Eqs.(\ref{gb}) and (\ref{He1}) we find
		\begin{equation}
			-g_{tt}(x_B)=\left(\frac{x_B-1}{x_B+1}\right)\left(1+\dfrac{c_1}{x_B-1}\right)+ \mathcal{O}(x_B-1)^2~,
		\end{equation}
		which from the BC on the deformed surface $ g_{tt}(1+\delta x_B)=0$, we obtain $c_1=1-x_B$, as noted in Table \ref{tb1}.

	\end{subequations}
	\begin{subequations}
		\renewcommand{\theequation}{\thesection.\arabic{equation}}
		\section{Comparison with the Quantum Love number}
		
		We wish to compare the classical and quantum calculations of the Love number of the QBH.

		First, let us emphasize the striking similarity between the expression for the classical Love number Eq.~(\ref{k2fluid}) and the expression in Eq.~(\ref{k22}) for the quantum Love number.
		If one identifies the expectation values with the corresponding classical observables $\langle \Psi_0|\widehat{Q}_{ij}|n,2,0\rangle=Q_n$ and the internal excited energy spectrum $|\Delta E_{1,n}|=\Delta E_{n}^{\text{int}}$, Eq.~(\ref{k2fluid}) and Eq.~(\ref{k22}) become identical.  As discussed in \cite{Brustein:2020tpg}, the quadrupole matrix element is given by the following integral:
		\begin{gather}
			|\langle \Psi_0|\widehat{Q}|n,2,0 \rangle|\!\!~=~\!\! \int d^3r   \delta\tilde{\rho}_{n,2}(r) r^2\;  Y_{20}\; \Psi_{n,2}~,
		\end{gather}
		where $\delta\tilde{\rho}_{n,2}(r)$ and $\Psi_{n,2}$ are the effective energy density and the mode function of the $n'$th excited level, respectively. In comparison to the overlap integral Eq.~(\ref{oi}), we find that the two expressions coincide when the density profile is decomposed by $\delta \rho=\sum\limits_n\delta \tilde{\rho}_{n,2} \Psi_{n,2}$ \cite{Chakrabarti:2013xza}.

		As anticipated in \cite{Brustein:2020tpg}, this correspondence is an explicit manifestation of the Bohr correspondence principle, which states that for macroscopic states associated with large quantum occupation numbers, expectation values correspond to classical quantities. The states that we consider are indeed states with large occupation numbers since for a given quantum state whose energy scales as $M \omega_n^2 R^2$, the occupation number $N$ scales as $N\hbar \omega_n\sim M \omega_n^2 R^2$, so $N\sim (\omega_n R)M R/\hbar\sim (\omega_n R)S_{{BH}}\gg 1$. Here $S_{BH}$ is the Bekenstein-Hawking entropy of the QBH.
		
		Equation~(\ref{finalc1}) completes the comparison of the quantum calculation to the classical calculation by specifying the required additional BC for the perturbation equations that determine the external metric.

		\section{Summary and Conclusions}

		We showed how to calculate $k_2$ in an explicit way from partial knowledge of the internal spectrum of an ultracompact object, be it the spectrum of the nonrelativistic fluid modes of a classical ultracompact star or the spectrum of excited states of a QBH. The single additional BC that encodes the relevant information about the interior of the QBH is determined in terms of its spectrum.
		In both cases, $k_2$ depends most strongly on the first excited level, or on the lowest lying fluid mode, and is proportional to the relative excitation energy of this level $k_2 \sim \Delta E/M$. The proportionality coefficient depends on additional information: the order unity ratio of the quadrupole moment of the excited level to its excitation energy $\Delta E$.
		
		Furthermore, since finding $k_2$ is equivalent to finding the ratio of two numbers, one does not need the detailed solution in the interior. The interior information is accessible to an external observer only through deformations of the surface of the QBH and can be integrated out, such that the only relevant quantity is a single boundary condition.
		
		That $k_2$ does not vanish for a QBH is a violation of the no-hair property and reflects the main difference between BHs to QBHs. For classical BHs, geometric deformations of the BH do not affect the asymptotic moments while for a QBH they do, and require a
		physical matter deformation, or equivalently, a physical response of the state of the QBH to the external perturbation and can therefore be detected by an external observer. Our results further highlight the importance of $k_2$ as a key diagnostic observable for probing the quantum nature of BHs.

		The agreement between the classical and quantum calculations of $k_2$ indicates that they are equivalent ways of deriving the same answer and strengthens the validity of each of the calculations.   The consistency of the classical and quantum cacluations provides us with an explicit  dictionary, translating quantum observables into classical GR quantities. Moreover, it demonstrates that the relation between quantum observables and the analogous GR quantities is in complete agreement with the Bohr correspondence principle.
		
		For completeness, we wish to emphasize the differences between the Love number of QBHs to that of the semiclassical objects (like gravastars and wormholes) reviewed in \cite{Cardoso:2017cfl}.
		First, these objects are not quantum in nature, they do not possess an event horizon and their unperturbed surface lies a finite distance away from the would-be horizon rather than at $R = 2M$. Second, the method for calculating their Love number is rather different; one must assume that a fictitious infinitely thin rigid shell surrounds these objects. The shell is made of a fictional matter violating both the weak and dominant energy conditions. The presence of a thin shell is necessary to guarantee the continuity of the interior to the exterior solutions. As a result, in order to compensate for the metric discontinuity, the boundary condition imposed on the object's surface is the so-called Israel junctions condition, which is different than the regularity condition of the external metric (see Table.~\ref{tb1}).
		Therefore, the origin of the Love number is in the discontinuity of the metric solution, which is a purely geometric property. On the other hand,
		the QBH Love number originates in the deformation of the interior matter distribution or equivalently in the coupling of its ground state to higher states by the external tidal perturbation.

		The current work can be extended to the spinning case, for which, as pointed out in \cite{Brustein:2020tpg}, spin effects are subleading and induce small corrections to $k_2$. Additionally, the spectrum of internal modes is expected to determine also the spectrum of ringdown modes and the nature of the merger. Thus, the internal structure of the QBH could potentially induce significant modifications to the merger and the ringdown phases of binary BH coalescence events; in particular, if resonance excitations of the QBH occur during the inspiral phase. We hope to report on these interesting possibilities  and their imprint on the emitted GW waveform in a future publication \cite{reso}.

	\end{subequations}
	\section*{Acknowledgments}
	The research of R. B. and Y. S. was supported by the Israel Science Foundation Grant No. 1294/16. The research of Y. S. was supported by the Negev scholarship.


\end{document}